# CURRENT ASSISTED, THERMALLY ACTIVATED FLUX LIBERATION IN ULTRATHIN NANOPATTERNED NbN SUPERCONDUCTING MEANDER STRUCTURES

H. Bartolf, A. Engel, A. Schilling

*Physics-Institute, University of Zürich, Winterthurerstrasse 190, 8057 Zürich, Switzerland*

K. Il'in, M. Siegel

*Institute of Micro- und Nano-electronic Systems, University of Karlsruhe, Hertzstrasse 16, D-76187 Karlsruhe, Germany*

H.-W. Hübers and A. Semenov

*DLR Institute of Planetary Research, Rutherfordstrasse 2, 12489 Berlin, Germany*

**Abstract**

We present results from an extensive study of fluctuation phenomena in superconducting nanowires made from sputtered NbN. Nanoscale wires were fabricated in form of a meander and operated at a constant temperature $T \approx 0.4 T_c(0)$. The superconducting state is driven close to the electronic phase transition by a high bias current near the critical one. Fluctuations of sufficient strength temporarily drive a section of the meander structure into the normal conducting state, which can be registered as a voltage pulse of nanosecond duration. We considered three different models (vortex-antivortex pairs, vortex edge barriers and phase slip centers) to explain the experimental data. Only thermally excited vortices, either via unbinding of vortex-antivortex pairs or vortices overcoming the edge barrier, lead to a satisfactory and consistent description for all measurements.

I.  INTRODUCTION

Thermodynamic fluctuations in superconductors have been studied for many decades, because they influence all properties of a superconductor, e.g. conductivity, susceptibility and specific heat [1], and allow deep insights into very basic aspects of the underlying physics [2-4]. In particular in one- and two-dimensional systems thermodynamic fluctuations play an important role leading to a rounding of the electronic phase transition even in very clean and homogeneous samples. The conductivity of one- and two-dimensional superconductors is particularly well understood, and detailed theories exist allowing an accurate description of many observed effects. Examples are the paraconductivity above the critical temperature $T_c(0)$ with distinct temperature dependences according to the dimensionality of the system [5-7], the non-vanishing resistance below $T_c(0)$ described by a Berezinskii-Kosterlitz-Thouless (BKT) phase transition in 2D-films [8-11], or the existence of phase-slip centers in 1D-wires [12-15].

Thermodynamic fluctuations are most easily observed near the phase transition. Therefore most experimental studies are done in a temperature range close to $T_c(0)$. At lower temperatures the probability of thermodynamic fluctuations drops exponentially so that they are experimentally no longer observable far below the transition temperature. However, the freezing-out of thermal fluctuations opens up the possibility to observe quantum fluctuations that prevail in the limit $T = 0$, for example quantum phase-slips [16].

Although well-defined one- and two-dimensional systems have been studied in great detail, the cross-over region between these limiting cases is less understood. This situation is just beginning to change as the size of superconducting conduction paths of devices such as SQUIDs or quantum detectors is continually decreasing, and therefore a better understanding of superconducting structures that are in between the limiting dimensions is required.

In one-dimensional wires, where both transverse dimensions are of the order or smaller than the shortest relevant length scale (i.e., the coherence length $\xi$) and only the longitudinal dimension in the direction of an applied bias current $I_b$ is much larger, fluctuations can stimulate phase-slip centers, either thermally or quantum-mechanically. Increasing one transverse direction, the width $w$, and keeping the thickness $d$ small, one opens up new excitation possibilities. It has been shown [17] that magnetic vortices, in two-dimensional superconductors, can exist as soon as $w \geq 4.4\xi$. However, before magnetic vortices can enter such a superconducting strip, either due to an externally applied magnetic field or to the magnetic self-field caused by $I_b$, they have to overcome an edge barrier [18] similar to the Bean-Livingston barrier [19] for three-dimensional, macroscopic superconductors. Interestingly, the critical width of about $4.4\xi$ for the cross-over from one- to two-dimensional behavior has been recently confirmed by a numerical comparison between the excitation energies for phase-slip centers and vortex excitations over the edge barrier [20].

In this paper we study the effect of thermal fluctuations in long superconducting NbN meanders with strip widths ranging from about 12 to 38 times the coherence length. These kinds of structures are used to realize superconducting nanowire single-photon detectors [21]. They are operated at an $I_b$ close to the experimental critical current $I_{c,e}$ and are sensitive in the visible and near-infrared spectral range (3.1 - 0.4 eV). It is generally believed that fluctuations are the major source of dark-count events in these detectors [22-24]. Measuring the dark-count rate thus gives us direct information about the fluctuation rates in a part of the superconducting phase diagram that is otherwise not easily accessible. The commonly used approach [23] to measure the DC resistance that is then used to infer the fluctuation rate is not appropriate at large bias currents close to the experimental critical current $I_{c,e}$, since the Joule heating cannot be eliminated. By contrast, Joule heating may influence the amplitude and duration of individual voltage transients in our time-resolved measurements, but it does not affect the count rate as long as it remains small compared to the reciprocal transient duration.

In the present study we have measured the dark-count rates for three meanders with equal thickness but different strip widths at a constant temperature of about $0.4T_c(0)$ as a function of the applied bias current. In the following chapter we will describe details of the sample preparation and the characterization of the superconducting state. In chapter III we present time-resolved measurements of fluctuation rates, and in chapter IV mathematical models are discussed where we consider the current-assisted thermal break-up of vortex-antivortex pairs, vortices overcoming the edge-barrier and phase-slip events as possible fluctuation origins.

## II. SAMPLE CHARACTERIZATION

### A. Nanoscale Sample Fabrication

The superconducting NbN films were deposited by dc reactive magnetron sputtering of a pure Nb target in an Ar+N$_2$ gas mixture, at a total pressure of about $10^{-3}$ mbar. The epipolished R-plane sapphire substrates were kept at 750°C during the film growth. The thickness $d$ of the resulting film was inferred from the sputtering time and a pre-determined deposition rate of 0.17 nm/s. The film growth was optimized with respect to the partial pressure of N$_2$ and the deposition rate to provide the highest transition temperature for the studied 6 nm high NbN films. More details about the film growth have been published previously [25].

The freshly deposited films were structured into a nanoscale meander by a combination of electron-beam lithography and reactive ion etching in a SF$_6$/Ar plasma discharge created by an applied radiofrequency power. The plasma was operated at low microwave power in order to minimize damages from ion bombardment and low pressure to ensure vertical bombardment during etching [26]. The three samples investigated had widths of the conduction path of 53 nm (sample 1), 83 nm (sample 2) and 171 nm (sample 3), respectively.

In Fig. 1 we show an electrograph of a clone of sample 1 having a filling factor *FF* of 16%. The filling factor is defined as the fraction the conduction path covers with respect to the whole area of the structure. The geometrical dimensions of the meanders were determined from such pictures taken with a scanning electron microscope (SEM). The samples for the measurements were fabricated with exactly the same process parameters as the meanders for the SEM characterization, therefore we expect them to show the same path width within an uncertainty of about 2 nm. The film height as expected from the sputtering rate multiplied by the deposition time was confirmed with an atomic force microscope (AFM).

After film growth and nanopatterning, the exposure to air of the NbN leads to an oxidization of surface and edge layers and to a suppression of superconductivity within these layers, which will influence the superconducting core of the conduction paths via the superconducting proximity effect [27]. Therefore, we used reduced values for the quantitative analysis of the experimental data, i.e. we subtracted 5 nm from the width determined with the SEM and 1 nm from the height determined with the AFM.

Electrical connections to the bond pads were made using contact photolithographic methods. The bond pads were connected to the printed circuit board of the measurement setup using a wedge wire bonder. Our layout allows for four probe electronic transport measurements, which is important for a reliable determination of material parameters as discussed in the following.

In principle, our versatile GDSII-design allows for the generation of more than 1000 devices with individual nanoscale dimensions on a 2" wafer and to interconnect them electronically [28].

B.   Resistivity Measurements and Electronic Parameters

The meander structures used in this study have been characterized with respect to their normal-state and superconducting properties. We will describe below the mathematical formalism that allows us to extract material parameters from the measured data. The thus obtained material parameters served as input parameters for the theoretical models describing the measured current-induced fluctuations discussed in the next chapter and to reduce the number of free parameters.

The transport measurements were done at low current $I_b \leq 500$ nA in a Physical Property Measurement System from *Quantum Design* in various magnetic fields $B$ up to 9 T perpendicular to the film surface. The temperature dependence of the experimental critical current $I_{c,e}(T)$ in zero magnetic field was obtained using a voltage criterion $V \leq 10$ mV, with a voltage of ~ 1 V in the normal conducting state. We calculated the square resistance $R_S(T) = R(T) \cdot \left( \frac{L}{w} + \frac{N}{2} \right)^{-1}$ of each structure, because it requires the SEM-measured input parameters which are less erroneous than the AFM-measured film height necessary for the calculation of the specific resistance. Here $L$ is the entire length of the nanowire and $N$ is the number of islands connecting the strips. For the critical current measurements, the respective I-V curves were taken at a fixed temperature.

A rather small electron mean free path $l$ [29] is characteristic for our NbN films, putting them into the dirty limit $l \ll \xi$, but still far from the metal-insulator transition as determined by the Ioffe-Regel criterion [30].

The electronic transition of nanoscale superconductors is usually rounded by thermal fluctuations on the order of $k_B T$ leading to paraconductive effects in the normal conducting state. The temperature dependence of the resistance data $R_S(T)$ reflects the dimensionality of the superconductor. The critical temperature in zero magnetic field $T_c(0)$ and the square resistance in the metallic state $R_{SN}$ were evaluated by least-squares fitting of the experimental data of $R_S(T)$ using a Cooper-pair fluctuation model developed for two-dimensional superconducting systems [5]

$$R_S(T) = \frac{R_{SN}}{1 + R_{SN} \cdot C \cdot \frac{1}{16} \frac{e^2}{\hbar} \frac{1}{t-1}}, \qquad t = \frac{T}{T_c(0)}, \qquad T > T_c(0) \qquad (1)$$

(see table 1, Fig. 2). Here $\hbar$ is the reduced Planck constant, $e$ is the elementary charge, and $C$ is a fitting parameter with $C = 1$ in an ideal system. We used the resistance data in the range $0.9 R_{SN} \leq R_S(T) \leq R_{SN}$ ($t > 1.13$) as the input data for the fitting procedure. We have set $R_{SN} = R_S(22\,\text{K})$ as $R_S(T)$ only weakly depends on $T$ right above the transition region. The resulting fit describes the measured resistance data very well down to about $0.5\,R_{SN}$ or $t > 1.014$, respectively. The fitting parameter $C$ turned out to be $\sim 2$ for all samples investigated.

Fitting analogous theoretical predictions for one-dimensional and three-dimensional superconductors [31] to our data, however, resulted in a deviation of the fit from the experimental data immediately outside the data interval used for the fitting, and the constant $C$ deviated about one order of magnitude from its ideal value.

Therefore we conclude that even the structure with the narrowest conduction path is indeed a 2D system as one would expect from the ratio of $d/\xi(0)$ and $w/\xi(0)$ (see table 1).

Further support for the two-dimensional character of our films comes from the inset of Fig. 2. Here $\left(R_S^{-1}(T) - R_{SN}^{-1}\right)^{-1}$ is plotted as a function of $T$. The linear $T$-dependence of the data well above $T_c(0)$ is only compatible with a 2D fluctuation conductivity formula described by Eq. (1), and incompatible with its one-dimensional and three-dimensional counterparts.

The critical temperatures $T_c(B)$ and the upper critical fields $B_{c2}$ were determined using a 50% resistance criterion. This criterion allowed us to determine the $T$ dependent magnetically induced phase transition line $B_{c2}(T) = \mu_0 H_{c2}(T)$, with $\mu_0$ the permeability of the vacuum. As shown in Fig. 3, the upper critical field is to a very good approximation linear in temperature, as expected from Ginzburg-Landau (GL) theory in the vicinity of $T_c(0)$.

The dark-counts triggered by thermal fluctuations were measured at a constant temperature of $\approx 0.4 T_c(0)$ which is well below the validity range of the usual GL [32, 33] approximations. To avoid the use of the exact, but complicated $T$-dependences of physical quantities that can be derived from microscopic theories [34, 35], we have used simple analytical expressions to approximate the real temperature dependences of these quantities as described below.

From the slope of the upper critical field $B_{c2}(T)$ close to $T_c(0)$, we determined the diffusivity of the quasiparticles in the normal conducting state (see also Appendix in [29]),

$$D = -\frac{4 \cdot k_B}{\pi \cdot e} \cdot \left(\frac{dB_{c2}(T)}{dT}\right)^{-1}_{T=T_c(0)}, \tag{2}$$

with $k_B$ the Boltzmann constant. Using the Einstein relation [36, 37]

$$DOS(E_F) = \frac{1}{(e^2 \cdot \rho_N \cdot D)} \tag{3}$$

we determined the total electronic density of states at the Fermi level, where $\rho_N = R_{SN} d$ is the normal state resistivity. This quantity is directly related to Sommerfeld's constant $\gamma_e = 1/3 \pi^2 k_B^2 DOS(E_F) = 220 \, \text{J/K}^2\text{m}^3$ (averaged over all three samples), which is consistent with the specific heat data obtained on standard NbN films used for thermal detector applications [38-41].

Within the framework of the Ginzburg-Landau (GL) theory, $B_{c2}(T)$ is related to the magnetic flux quantum, $\Phi_0 = h/2e \approx 2.07 \cdot 10^{-15}$ Vs and to the temperature dependent coherence length $\xi(T)$ by

$$B_{c2}(T) = \frac{\Phi_0}{2\pi \, \xi(T)^2}. \tag{4}$$

From a linear extrapolation of $B_{c2}(T)$ to zero temperature one obtains the zero temperature GL coherence length $\xi_{GL}(0)$. It was shown [42-44] that a realistic value for the true upper critical field $B_{c2}(0)$ of dirty superconductors can be obtained by multiplying the extrapolated value with 0.69. In order to model the full temperature dependence of the true coherence length $\xi(T)$ we used an analytical formula that approximates Werthammers $T = 0$ result and includes the GL dependence near $T_c(0)$

$$\xi^2(t) = \xi^2(0) \cdot (1-t)^{-1} \cdot (1+t)^{-0.5}, \qquad \xi^2(0) = \sqrt{2}\xi_{GL}^2(0). \tag{5}$$

NbN is a strongly-coupled superconductor. Therefore, we express the zero temperature energy gap by the experimentally determined relation $\Delta(0) = 2.08 \cdot k_B \cdot T_c(0)$ [45]. The temperature dependence of the gap is modeled by the approximate formula [46]

$$\frac{\Delta(t)}{\Delta(0)} = \tanh\left(1.82 \cdot \left[1.018\left(\frac{1}{t}-1\right)\right]^{0.51}\right). \tag{6}$$

This formula virtually coincides with the numerical values of Mühlschlegel [35].

The magnetic penetration depth $\lambda(0)$ can be expressed in the dirty limit as $\lambda(0) = (\hbar\rho_N / \pi\mu_0\Delta(0))^{0.5}$. Its temperature dependence is given by (Eq. 3.134 in [33])

$$\frac{\lambda(T)}{\lambda(0)} = \left[\frac{\Delta(T)}{\Delta(0)} \cdot \tanh\left(\frac{\Delta(T)}{2k_BT}\right)\right]^{-0.5}. \tag{7}$$

In order to keep the model to be explained in chapter IV as simple as possible, we use instead an analytic approximation for the temperature dependence of the penetration depth,

$$\frac{\lambda(t)}{\lambda(0)} = (1-t^2)^{-0.5}(1+t^{1.5})^{-0.25}. \tag{8}$$

For practical purposes Eqs. (7) and (8) again virtually coincide. The effective penetration depth in thin films with $d \ll \lambda$ [47] is then

$$\Lambda(T) = \frac{2 \cdot \lambda^2(T)}{d}. \tag{9}$$

This quantity is about 170 times larger than $\lambda$ for bulk NbN and exceeds our samples' dimensions for all temperatures. Therefore the current-density distribution can be assumed to be homogeneous at all temperatures, allowing for a straightforward determination of the depairing critical-current from current-voltage curves. In order to keep the explicit dependence on the order parameter of the GL model, we used the following expression for the depairing critical-current in the dirty limit [48]

$$I_{c,d}(t) = \frac{2 \cdot \sqrt{2\pi} \cdot (e^\gamma)^2}{21 \cdot \zeta(3) \cdot \sqrt{3}} \cdot \frac{[\Delta(0)]^2}{\sqrt{k_B \cdot T_c(0)}} \cdot \frac{w \cdot d}{e \cdot \rho_N \cdot \sqrt{D \cdot \hbar}} \cdot (1-t^2)^{3/2} \cdot (1-t^4)^{1/2}, \quad (10)$$

which reduces to the conventional form (two-fluid model [49]) on substituting the BCS relation between $T_c(0)$ and $\Delta(0)$. Here $\gamma = 0.577$ is Euler's constant and $\zeta(3) = 1.202$ is Apery's constant.

Our experimental critical currents $I_{c,e}$ at $T \approx 0.4 T_C(0)$ are about 60% of the theoretical limit computed using Eq. (10). This is an indication of the excellent uniformity of the conduction paths of the present meander structures. Already a small number of constrictions along the total length of the meander would significantly limit the experimental critical current, whereas the material parameters entering Eq. (10) as determined from resistivity measurements are not noticeably affected by small variations of the cross-sectional area.

From the above analysis of the NbN meander structures we conclude that our NbN films belong to the class of two-dimensional, strongly-coupled type-II superconductors with a high square resistance $R_{SN}$, which is only about one order of magnitude smaller than the resistance quantum $h/2e^2 \cong 13 \, \text{k}\Omega$. Therefore, one can expect to observe a pronounced transition of the BKT-type [8-11].

In thin films of this type it is very probable that thermal fluctuations excite pairs of vortices. These pairs consist of single vortices with the respective supercurrents circulating in opposite directions and leading to a bound state called a vortex-antivortex-pair (VAP). The necessary criterion for a BKT transition, where the VAPs are starting to dissociate, is a logarithmic dependence of the vortex interaction

$$U(r) = A(T) \cdot \ln\left(\frac{r}{\xi(T)}\right) + 2\mu_{\text{C}}(T), \qquad A(T) = \frac{\Phi_0^2}{\pi \cdot \mu_0 \cdot \Lambda(T)} \qquad (11)$$

from the distance of the vortex core centers $r \geq \xi(T)$ [50]. Here $A(T)$ is the vortex interaction constant and $\mu_{\text{C}}(T)$ the vortex core potential which is defined as half the free energy of a pair at smallest separation. The necessary condition for a BKT transition is fulfilled in our nanoscale NbN meanders because the effective penetration depth is much larger than the sample's dimensions. Below the ordering temperature $T_{\text{BKT}}$, all VAPs are bound. Above $T_{\text{BKT}}$ VAPs break up by thermal fluctuations, and single vortices and VAPs coexist in thermodynamic equilibrium. Therefore, if pinning is neglected, the dissociated VAPs may move due to the Lorentz-force exerted by the probing current $I_{\text{b}}$ and cause a finite resistance in the temperature region $T_{\text{BKT}} < T < T_{\text{c}}(0)$ according to [50]

$$\rho(T) = a \cdot \exp\left(-2 \cdot \sqrt{b \cdot \frac{T_{\text{c}}(0) - T}{T - T_{\text{BKT}}}}\right), \qquad (12)$$

with $a$ and $b$ material dependent parameters used for fitting our data. In Fig. 2, we show a least-squares fit to the experimental data in this temperature interval from which we determined $T_{\text{BKT}}$ (see table 1). With the BMO-relation (shown by M. R. Beasley, J. E. Mooij and T. P. Orlando (BMO) [50, 51])

$$\frac{T_{\text{BKT}}}{T_{\text{c}}(0)} = \frac{1}{1 + 0.173 \cdot \frac{\varepsilon_{\text{BKT}}}{\pi} \cdot R_{\text{SN}} \cdot \frac{2e^2}{h}} \qquad (13)$$

we also determined the polarizability $\varepsilon_{\text{BKT}} \approx 10$ of a VAP at the BKT vortex phase transition in the presence of other VAPs, and successfully crosschecked it with the universal relation $k_{\text{B}} \cdot T_{\text{BKT}} = A(T_{\text{BKT}})/4\varepsilon_{\text{BKT}}$ for topological two-dimensional phase transitions, first shown by Nelson and Kosterlitz [50, 52].

In Fig. 2 we show that the temperature dependence of $\rho(T)$ follows qualitatively the behavior that one would expect for finite-size BKT-systems [50]. In such systems thermally unbound VAPs can exist even below $T_{BKT}$ and lead to a resistive tail for $T \to 0$ [53].

In table 1 we have summarized the material parameters from our analysis of the resistivity data for the three samples investigated.

### III. TIME-RESOLVED DETECTION OF FLUCTUATIONS

#### A. Pulse-Detection Setup

To measure time-resolved fluctuation effects the meanders were thermally anchored to the cold plate of a $^4$He-bath cryostat. The ambient temperature of the meander holder was $T \cong 5.5 \text{ K}$ under operating conditions. As the superconducting meanders are efficient single-photon detectors by design, they were shielded against photons from blackbody radiation with an Al foil that was also thermally connected to the cold plate.

A bias current $I_b$ was supplied by a custom-made, battery powered constant-voltage source to ensure a low-noise. The critical currents listed in table 1 were measured using this configuration. The bias current passed two low-pass filters. One located at the top of the cryostat at room temperature, and the other one on a printed circuit board near the holder, where it passed a bias-tee and finally the nanoscale meander structure. With an increase of $I_b$ near the critical current, the meanders approach the transition into the normal conducting state and become particularly sensitive to fluctuations or to an externally deposited energy of any kind.

The fluctuation energy inside the meanders acts as a seed for the formation of a normal conducting domain which again disappears on a timescale of a few hundred picoseconds [54]. Due to the finite kinetic inductance of the system the resulting voltage pulses decay on a time scale that is larger than the lifetime of the normal-conducting domains [55]. These pulses were passed to a high electron-mobility transistor microwave-amplifier chain with an effective band width of 0.1-2 GHz and a total gain of 48 dB. The amplified voltage signals were then fed with a 50 Ω coaxial cable into the readout electronics, that consisted of either a 6 GHz bandwidth single-shot digital oscilloscope (Wave Master 8600A from *LeCroy*), or a 300 MHz bandwidth gated voltage-level threshold counter (SR400 from *Stanford Research Systems*).

B.     Measured Fluctuation Rates

In Fig. 4 we present the measured fluctuation rates $\Gamma$ as a function of $I_b$ for the structures characterized by the transport measurements as described in the previous chapter. The data were modeled with the parameters of table 2 according to the scenarios described in the next chapter. Close to the experimental critical current $I_{c,e}$ ($I_b / I_{c,e} \approx 1$) the fluctuation rates increase virtually linearly on a logarithmic scale as $I_b$ approaches $I_{c,e}$ from below. At lower bias currents, where the measured rates dropped to about 10 to 1 Hz, the measured data deviate significantly from the approximately exponential behavior, however. The narrow structures with $w < 100\,\text{nm}$ (samples 1 and 2) exhibit a tail-like structure below $I_b / I_{c,e} \approx 0.88$, with higher than expected fluctuation rates. As we will discuss below, these data can be quantitatively explained by the Lorentz-force induced motion of single, unbound vortices and antivortices due to finite size effects [53] within the BKT model (see chapter IV. A.). In the wider structure, by contrast, the fluctuation rates decrease even faster than expected by a simple exponential dependence which we cannot explain within the formalism of chapter IV. However, we cannot exclude the possibility that these deviations from the exponential behavior below around 1 Hz originate from our electronic setup.

The lines drawn in Fig. 4 result from a fit to two established theories that we will discuss in the next chapter, together with an additional scenario invoking phase slip phenomena.

## IV. FLUCTUATION MODELS

For modeling the measured fluctuation rates we explicitly include a current-dependence of all relevant parameters characterizing the superconducting state as they depend themselves on the current-dependent energy gap. In a previous work [54], we precisely determined the bias-current dependence of the kinetic inductance for structures similar to those investigated in this study. The kinetic inductance $L_{\text{kin}} = \mu_0 \lambda^2 L / dw$ of a superconducting device is directly related to the magnetic penetration depth, which in turn depends on the energy gap (see Eq. (7)). A reasonable approximation for the measured current dependent kinetic inductance

$$\delta(I_{\text{b}}) \equiv \left[\frac{\lambda(T, I_{\text{b}})}{\lambda(T, 0)}\right]^2 = \left(1 - 0.31 \cdot \left(\frac{I_{\text{b}}}{I_{\text{c,e}}(T)}\right)^{5/2}\right)^{-1/3} \tag{14}$$

suggests a 13% increase of the undisturbed penetration depth at the experimental critical current.

### A. Unbinding of Vortex-Antivortex Pairs

As outlined in chapter II the logarithmic dependence of the vortex interaction (see Eq. (11)) on the distance $r$ of the vortex-core centers is responsible for the topological breakdown of the ordered state in the vortex system at $T > T_{\text{BKT}}$.

The presence of VAPs could in turn also have a profound effect on the dynamics of the fluctuations, i.e. the fluctuation rate. In the following, for simplicity we will first neglect any finite size effects (i.e., $w \gg \Lambda$). Under this assumption all thermally excited vortices and antivortices will be paired at the operating temperature $T \cong 5.5$ K $< T_{\text{BKT}}$. The application of a bias current $I_{\text{b}}$ then exerts a Lorentz-force that is directed in opposite directions for the vortex and the antivortex, respectively. The resulting torque leads to an orientation of the VAPs perpendicular to the current direction, and the pair is pulled apart, thereby reducing the energy of the VAP. However, the current cannot pull apart the constituents to infinite distance, since the VAP self-energy grows logarithmically with the separation $r$ of its constituents. Consequently, by increasing the bias current, the VAP pass through a minimum binding energy (at $r = 2.6 \cdot \xi(T) \cdot I_{c,e}(T)/I_b$ according to [50]) that can be calculated in a straightforward mathematical variational calculus and reads

$$U_{VAP,m}(T, I_b) = \frac{A(T, I_b)}{\varepsilon} \cdot \left[ \ln\left(\frac{2.6 \cdot I_{c,e}(T)}{I_b}\right) - 1 + \frac{I_b}{2.6 \cdot I_{c,e}(T)} \right], \qquad (15)$$

where $\varepsilon$ is the averaged polarizability of a VAP within the entire VAP population [50]. This binding energy may be overcome by a thermal excitation with a probability equal to Boltzmann's factor $\exp(-U_{VAP,m}/k_B T)$. We restrict the following discussion to this minimum binding energy. Pairs with smaller or larger elongation will be unbound with respectively lower probability.

Because of the very high bias currents we may also neglect vortex-pinning effects. The thermally unbound vortices will thus move freely towards opposite edges of the strip where they will leave the structure or annihilate with an oppositely orientated vortex. In either case, the moving vortices will dissipate energy which initiates the creation of a normal conducting domain. Such domains cause voltage transients that are then registered as dark-count events. It is straightforward to assume that the resulting corresponding dark-count rate is proportional to the unbinding probability, and therefore

$$\Gamma_{VAP,m}(T, I_b) = \alpha_{VAP} \cdot \exp(-U_{VAP,m}(T, I_b)/k_B T), \qquad (16)$$

with $\alpha_{VAP}$ a proportionality constant with the meaning of an attempt rate.

In the meanders with strip widths $w \ll \Lambda$ VAPs can be thermally unbound even in the absence of a bias current (see explanation of the resistive tail of Fig. 2 in chapter II. B.). The corresponding density of unbound vortices and antivortices will add a background of dark-count events since they will also start to move under the action of the Lorentz-force. The

density of free vortices in a strip with width $w \ll \Lambda$ can be derived following [50, 53, 56] as

$$n_{\mathrm{SV}} = \frac{1}{\xi(T)^2} \exp\left(-\frac{A(T,I_b)}{k_B T}\left[\frac{1}{\gamma_0} + \frac{l_w}{2\varepsilon}\right]\right), \qquad l_w = \ln\left(\frac{w}{\xi(T)}\right), \qquad (17)$$

and the background dark-count rate stemming from these unbound vortices should be proportional to $n_{\mathrm{SV}}$ according to

$$\Gamma_{\mathrm{SV}}(I_b, T) = \alpha_{\mathrm{SV}} \cdot n_{\mathrm{SV}}, \qquad (18)$$

with a proportionality constant $\alpha_{\mathrm{SV}}$ and again with a current dependence stemming from the current dependence of the vortex interaction constant $A$.

Eq. (17), contains the vortex core energy

$$\mu_C(T, I_b) = \frac{A(T, I_b)}{\gamma_0}, \qquad (19)$$

where $\gamma_0$ is a parameter of order unity.

In a London-type model of a vortex, its core is approximated by a metallic cylinder with radius $\xi$. Using relative simple algebra, one finds that the associated loss of condensation energy inside the London vortex volume sets an upper limit for $\gamma_0 \leq 8$ [50]. However, the dotted lines in Fig. 4 were obtained by setting $\gamma_0 = 4$, $\varepsilon = 2.5$ and $\varepsilon = 3.2$ for sample 1 and sample 2, respectively.

B.   Vortices overcoming the edge barrier

At bias currents close to the depairing-critical current the magnetic self-field at the strip edges is much larger than the lower critical field $B_{c1}$ for vortex entry. The entry of vortices at one edge and antivortices at the opposite edge is only prohibited by an edge barrier very similar to the Bean-Livingston surface barrier [19]. The existence of such a barrier results from the requirement of vanishing components of the supercurrents encircling the vortex core that are parallel to the current path. In the case of narrow strips, this condition may be formally fulfilled by introducing an infinite chain of mirror vortices and antivortices at both edges [18].

In this picture, the interaction of the magnetic moment of the vortex with the supercurrents of the virtual antivortices leads to an effective attractive force towards the near edge of the superconductor. The resulting Gibbs free energy in the London-limit, neglecting the finite size of the vortex core, including the potential from a bias current for a single vortex at a distance $x$ from the edge and is given by [18, 57, 58]

$$G(T, I_b, x) = E_B(T, I_b) \cdot \left[ \ln\left( \frac{2w}{\pi \cdot \xi(T)} \cdot \sin\left( \frac{\pi \cdot x}{w} \right) \right) - \frac{I_b}{I_B(T, I_b)} \cdot \frac{\pi}{w} \cdot \left( x - \frac{\xi(T)}{2} \right) \right], \quad (20)$$

with

$$I_B(T, I_b) = \frac{\Phi_0}{2 \cdot \mu_0 \cdot \Lambda(T, I_b)} \quad \text{and} \quad E_B(T, I_b) = \frac{\Phi_0^2}{2 \cdot \pi \cdot \mu_0 \cdot \Lambda(T, I_b)} \quad (21)$$

being current and energy scales, respectively. These quantities are bias-current dependent through their dependence on the effective penetration depth. The first term of Eq. (20) corresponds to the nucleation energy of the vortex, while the second term describes its interaction with the bias current. An analogous potential can be derived for an antivortex entering from the opposite edge. An additional term describing the contribution due to an external magnetic field has been neglected here. The earth magnetic field, which was not shielded during our experiments, and the magnetic fields from neighboring strips are of the same order of magnitude, and their contribution is negligible compared to the other two contributions included in Eq. (20). The finite size of the vortex core is neglected in the derivation of Eq. (20), which leads to a diverging free energy for $x \to 0$ and $x \to w$. One has to expect that the real potential a vortex experiences deviates from the predictions of Eq. (20), when it comes to within a distance of about $\xi$ to the edge [59]. Due to the lack of detailed theories of real vortex penetration into the strip, we arbitrarily set the vortex' energy to zero at the position $x = \xi(T)/2$. As will be discussed later, the exact position for which the Gibbs free energy is normalized to zero does not change the physical picture.

The Gibb's free energy as described by Eq. (20) is plotted in Fig. 5 for all three samples in the absence of a bias current, and in addition for sample 1 for five different equidistant bias currents between zero and $I_{c,e}$. It can be clearly seen that there is an energy barrier for the vortex entry in all cases. However, the height of this barrier as well as its width shrinks with increasing bias current. It is a straightforward analytical problem to determine the barrier height as a function of temperature and bias current (see Table 2):

$$G_{B,max}(T,I_b) = E_B(T,I_b) \cdot \left[ \ln\left( \frac{2w}{\pi \cdot \xi(T)} \cdot \frac{1}{\sqrt{1+\left(\frac{I_b}{I_B(T,I_b)}\right)^2}} \right) - \frac{I_b}{I_B(T,I_b)} \cdot \left[ \arctan\left(\frac{I_B(T,I_b)}{I_b}\right) - \frac{\pi \cdot \xi(T)}{2w} \right] \right]$$
(22)

The corresponding probability for thermally activated hopping of a vortex over this energy barrier is proportional to the Boltzmann factor of the barrier maximum. Once a vortex jumped over the barrier it will move across the strip, agitated by the Lorentz-force term of the potential. Like in the VAP scenario, the motion of the vortices across the strip creates a normal conducting domain and finally results in a voltage transient. Because the magnetic self-field at the edges is proportional to the bias current, we assume the attempt rate to be linear in the bias current to a first approximation, and arrive at an expression for the resulting dark-count rate originating from thermally activated vortex hopping as

$$\Gamma_{VH}(I_b,T) = \alpha_{VH} I_b \exp(-G_{B,max}(T,I_b)/k_B T), \tag{23}$$

with the constant $\alpha_{VH}$ including the attempt rate and the details of the geometry.

In the low temperature limit $T \to 0$ the probability for thermal fluctuations freeze-out exponentially and give way to macroscopic quantum mechanical tunneling [60]. In this scenario vortices can enter the superconducting strip by tunneling through the barrier of Eq.

(20). According to Tafuri et al. [61] the probability for such an event is $\exp\left(-\beta_B \frac{\eta \cdot x_B^2}{\hbar}\right)$. The barrier shape enters via the parameter $\beta_B$ of order unity, $\eta$ is the vortex drag coefficient within the Bardeen-Stephen model for vortex motion [62] and $x_B$ is the width of the barrier (see Fig. 5). The barrier width can be calculated numerically as the root of the Gibbs free energy potential Eq. (20). In order to obtain the tunneling rate of vortices, one has to multiply the tunneling probability with an attempt rate that we assume to be proportional to the bias current, for the same reasons as in the case of thermally activated vortex hopping. The dark-count rate can then be expressed as

$$\Gamma_{VT}(I_b, T) = \alpha_{VT} I_b \exp\left(-\beta_B \frac{\eta \cdot x_B^2}{\hbar}\right), \tag{24}$$

again with an attempt rate $\alpha_{VT}$.

### C. Phase Slip Centers

In small superconducting wires with cross-sectional area $A = wd$ on the order of $\xi^2$ fluctuations can lead to a nontrivial, temporary local destruction of the order parameter accompanied by a phase slip [12]. These phase slips can lead to a random succession of voltage pulses, possibly adding to the observed dark-count rate.

A more quantitative picture of thermal phase slips (TPS) was developed by Langer, Ambegaokar [13] and McCumber, Halperin [15] and later confirmed in experiments carried out on superconducting tin whiskers [63, 64]. McCumber [14] deduced the equivalence of this approach for current or voltage driven power sources.

The phase slip free energy barrier can be derived in terms of the condensation energy (equivalent to Eq. 8.4 in [33])

$$\Delta F_{PS}(T, I_b) = \frac{8\sqrt{2}}{3} \cdot \frac{B_c^2(T, I_b)}{2\mu_0} \cdot A \cdot \xi(T) . \tag{25}$$

Here $H_c$ is the thermodynamic critical field. Eq. (25) has a clear physical interpretation: The required energy is equal to the superconducting condensation energy contained in a volume of the cross-section of the wire, times the coherence length up to a numerical factor stemming from the variational derivation within the GL theoretical framework. We modeled the thermodynamic critical field by the following analytic formula

$$H_c(T, I_b) = H_c(0, I_b) \cdot (1 - t^2) \cdot (1 + t)^{-1/6} \tag{26}$$

that virtually coincides with the numerical data of Mühlschlegel [35]. Using this analytical temperature-dependence of the critical field and the temperature-dependence of the coherence length of Eq. (5), we can express the energy barrier for phase slip events as

$$\Delta F_{PS}(T, I_b) = 0.72 \cdot k_B T_c(0) \cdot \frac{h}{2e^2} \cdot \frac{1}{R_{SN}} \frac{w}{\xi(T)} \cdot (1-t) \cdot (1+t)^{7/6} \cdot (\delta(I_b))^2 \tag{27}$$

and express the parameters by more easily measurable quantities.

A resistive phase slip event can be interpreted as the following process: A driving voltage between two points that are connected by a superconductor will induce a steady growth of the absolute value for the phase $\varphi$ of the order parameter according to the Josephson rate [65]

$$\dot\varphi = \frac{2eV}{\hbar} . \tag{28}$$

In general, this requires a continuously increasing current [13]. However, if fluctuations deposit energy in the interior of the superconductor, a reduction of the phase difference (by a back-snap of the phase by $2\pi$) occurs at the same rate as the voltage increases, and a constant flow of charge can be maintained. The phase slip rate then adjusts to $\Gamma_{PS} = \frac{1}{2\pi}\dot\varphi$. The bias current makes phase slippage in the direction antiparallel to $I_b$ more probable than in the direction parallel to the current. The free energy difference between these two possibilities can be calculated to amount to $\delta F_{PS} = (h/2e) \cdot I_b$.

Using again Boltzmann statistics (equivalent to Eq. 8.7 in [33]), the thermally activated

phase-slip rate becomes

$$\Gamma_{\text{TPS}} = \alpha_{\text{TPS}} \cdot \exp\left(-\frac{\Delta F_{\text{PS}}(T, I_b)}{k_B T}\right) \cdot \sinh\left(-\frac{\delta F_{\text{PS}}}{2 k_B T}\right), \quad (29)$$

with the attempt frequency $\alpha_{\text{TPS}}$.

A quantum mechanism for the occurrence of such phase slips ("quantum phase slips", QPS) has also been suggested [66, 67]. One obtains a reasonable approximation for their probability by replacing the thermal energy scale $k_B T$ by the appropriate quantum mechanical scale $\hbar/\tau_{\text{GL}}$ in Eq. (29), where $\tau_{\text{GL}} = \pi \hbar / 8 k_B (T_c(0) - T)$ is the characteristic relaxation rate of non-equilibrium excitations in superconductors within the time-dependent Ginzburg-Landau theory. This gives us the rate for quantum mechanical phase slips [67]

$$\Gamma_{\text{QPS}} = \alpha_{\text{QPS}} \cdot \exp\left(-\frac{\Delta F_{\text{PS}}(T, I_b)}{\hbar/\tau_{\text{GL}}}\right) \cdot \sinh\left(-\frac{\delta F_{\text{PS}}}{2\hbar/\tau_{\text{GL}}}\right) \quad (30)$$

with the QPS attempt frequency $\alpha_{\text{QPS}}$.

D. Discussion

The models presented above are candidates to describe the measured fluctuation rates. The key quantity is always the necessary excitation energy $E_{\text{exc}}$ that enters the corresponding Boltzmann factor. In Fig. 6 we have plotted these energies for all three models as a function of the reduced bias current $I_b / I_{c,e}$ for sample 1 with the narrowest conduction path $w \approx 50$ nm, because the sample with the smallest spatial dimensions of this study should be most sensitive to fluctuations, either of thermal or of quantum mechanical nature. The excitation energies were calculated using Eqs. (15), (22), (27), the formalism introduced in the previous chapter and the parameters from table 1. In the relevant current range, i.e. $I_b / I_{c,e} > 0.7$, the excitation energies for the unbinding of VAPs and vortices hopping over the edge barrier are comparable to each other, whereas those for phase slip phenomena are significantly larger, thereby leading to an at least two orders of magnitude lower thermodynamic probability for the occurrence of such a fluctuation event. The results for the other samples are qualitatively similar. While the excitation energies for all models increase with increasing width, the values for phase slip events increase even faster as compared to the other mechanisms, which makes them even less likely to occur in wider samples.

Using the excitation energies from Fig. 6 we have calculated the fluctuation rates and plotted them on a logarithmic scale in Fig. 7 as a function of the reduced bias current. For each fluctuation model, the rates have been normalized to unity for $I_b / I_{c,e} = 1$, and the quantum tunneling of vortices through the edge barrier as well as quantum phase slips have been included. Again, the unbinding of VAPs and the hopping of vortices over the edge barrier are very similar to each other, but the other fluctuation mechanisms show a distinctively different current-dependence. If we compare these theoretical current-dependencies with the experimental data shown in Fig. 4, only the unbinding of VAPs and vortices hopping over the edge barrier exhibit a current dependence that is compatible with the experimental data.

Before we continue the discussion of these two promising fluctuation mechanisms, we want to briefly comment on why the other proposed models are not relevant for explaining the measured fluctuation rates. The lowest fluctuation rates come from the quantum tunneling of vortices through the edge barrier (see Fig. 7). One can define a cross-over temperature $T_{co}$ below which the probability for quantum tunneling becomes larger than the probability for a thermally activated nucleation and jump over the edge barrier by equating the exponential terms from Eqs. (22) and (23). The corresponding results for $I_b = 0$ are listed in Table 2. These cross-over temperatures are all well below 1 K, which is much lower than the operational temperature in our experiments. Tunneling of vortices in the present structures becomes therefore only relevant at sub-Kelvin temperatures.

The fact that thermal or quantum phase slips can be excluded as the cause of the observed fluctuation rates is a direct consequence of the relatively large strip width $w$ as compared to the coherence length $\xi$, which leads to high excitation energies. Even for sample 1, this width is by a factor of $\approx 10$ larger than $\xi(T)$. The physical validity of the phase-slip approach to the present situation has to be questioned in general, however, since it had been developed for superconducting wires, where both transverse dimensions are on the order of the coherence length or smaller. Nevertheless, it is very interesting to note that in case of phase slips, the rate of quantum-mediated fluctuations is *larger* than the thermally induced ones, just as the theory of QPS predicts [3]. Furthermore, we note that our results are in line with numerical calculations [20], showing that vortex-based thermal fluctuations should indeed dominate as long as Likharev's criterion $w \geq 4.4\xi$ is fulfilled.

We finally tried to fit the two thermally excited vortex models (*unbinding of VAPs* and *vortices hopping over the edge barrier*) to the experimental data in an attempt to distinguish between the two scenarios. In the vortex hopping model, we could independently determine all physical parameters, leaving only the attempt rate and the device temperature as free fit parameters. This procedure yielded quite satisfactory results of 5.7 K, 5.7 K and 4.5 K for the samples 1, 2 and 3, respectively. The temperature deviation of sample 3 might be explained by a better thermal isolation of the cryostat during this particular measurement. Another explanation might be a reduced barrier height of sample 1 and 2, which is equivalent to a higher model temperature. Such a reduction could be caused, e.g., by edge inhomogeneities or structural damage during device fabrication. Nevertheless as mentioned above, the physical validity of Eq. (20) is questionable at positions close to the edge of the strip. Accordingly, the height of the barrier and therefore the fluctuation rate depends on the position for which the Gibbs free energy of the vortex is set to zero. For example, setting $G(T, I_b, x = \xi(T)) = 0$ leads to model temperatures about one Kelvin less than the ones given above. The point for which the free energy of the vortex is zero near the strip edge might itself depend on the strip width.

For the thermal VAP unbinding scenario the device temperatures determined within the vortex-hopping model were used. Besides the attempt frequency $\alpha_{VAP}$ the polarizability of a VAP $\varepsilon$ remained as the only free parameter. Best fits were obtained for $\varepsilon \approx 1$, (see table 2). The apparent discrepancy between this value and the rather high $\varepsilon_{BKT}$ obtained from fitting the resistance data has its origin in different physical conditions for the corresponding experiments. The transport measurements were carried out at very low bias-currents, thus probing large VAPs with a high polarizability. The fluctuation rates, however, were measured at currents close to the critical current and therefore probing VAPs close to the minimum separation $r \approx 2.6 \cdot \xi(T)$ and hence with low polarizability $\varepsilon \approx 1$.

The best resulting fits according to both relevant models are plotted in Fig. 4. Within the accuracy of our data it is not possible to decide in favor of any of the two models. Based on the calculated excitation energies alone, we would expect that vortices are frequently thermally excited over the edge barrier. However, within the VAP unbinding model we can naturally explain the tail-like structure at relatively low currents observed in the fluctuation rates of the samples with the sub-100 nm wide conduction paths. Due to the small width of the conduction paths we can expect a finite density of thermally unbound VAPs even in the absence of a bias current. The estimated corresponding contribution to the fluctuation rate is also shown in Fig. 4 as dotted lines. For the sample with the widest conduction path, the density of thermally unbound vortices seems to be so low that the resulting fluctuation rate is beyond the sensitivity of the present experiment. This might explain the absence of the tail-like structure for sample 3. We want to mention, however, that this low-frequency tail in $\Gamma(I)$ is very sensitive to electronic noise in the measurement circuit. The present data could only be obtained after careful elimination of noise sources, and we can therefore not exclude that a further reduction of this noise would suppress the tail-like structures in Fig. 4 for the samples with the sub-100 nm wide conduction paths.

## V. CONCLUSIONS

We have presented an extensive study of current-induced fluctuation phenomena in superconducting nanoscaled meander structures at temperatures well below the superconducting transition. Fluctuation rates were studied as a function of the applied bias current in three samples with different strip widths. In such structures, fluctuations of sufficient energy lead to measurable voltage pulses with picosecond rise times and nanosecond duration that can be counted with a threshold-level voltage pulse counter.

Using established theoretical models and taking into account the experimentally determined current-dependence of the energy gap $\Delta$ we aimed to model our experimental data. At the operating temperature of $T \approx 5.5\,\text{K}$ and for bias currents $I_b / I_{c,e} \geq 0.7$, thermally activated or quantum mechanical phase slips as well as the quantum mechanical tunneling of vortices through the edge barrier, can be clearly excluded as the dominant mechanism leading to fluctuation-induced voltage transients.

The most likely explanation of the observed fluctuation rates involves thermally activated vortices moving across the strips, either as unbound vortex-antivortex pairs or as single vortices overcoming the edge barrier, but the available data do not allow us to finally decide in favor of one of the scenarios. A measurement of the temperature-dependence of the fluctuation rates might resolve this interesting question.

**Acknowledgment**

H.B. acknowledges support by the Swiss NCCR Manep. The nanostructuring was carried out at the FIRST Center for Micro- & Nanoscience of ETH Zürich. Throughout this paper the international SI unit system was used.

Table captions

Table 1. Material parameters of the investigated structures. The geometric dimensions were determined using SEM & AFM techniques. The formalism for determining the other material parameters is discussed in chapter II. For the modeling of the data, the width $w$ has been reduced by 5 nm and the thickness $d$ by 1 nm as justified in the text.

Table 2. Parameters for the fluctuation models, that were used to fit the measured dark count rates within the formalism developed in chapter III. In the case of vortex tunneling and both phase slip mechanisms, the given attempt rates $\alpha$ match the measured fluctuation rates at $I_b = I_{c,e}$. $T$ is the operation temperature assigned separately to each sample to fit the experimental data within the vortex hopping scenario (see text). The listed polarizabilities $\varepsilon$ provide the best fit to the measured data within the VAP scenario. The excitation energies and all the model parameters were calculated for zero bias current ($U_{\text{VAP,m}}/k_\text{B}$ for $I_b = 0.01 I_{c,e}$).

Table 1

| Sample # | $w$ | $L$ | $d$ | $N$ | $FF$ | $R_{SN}$ | $T_c(0)$ | $T_{BKT}$ | $\varepsilon_{BKT}$ | $I_{c,e}$ | $D$ | $DOS(E_F)$ | $\Delta(0)$ | $\xi(0)$ | $\lambda(0)$ | $\Lambda(0)$ |
|---|---|---|---|---|---|---|---|---|---|---|---|---|---|---|---|---|
| | nm | µm | nm | | | Ω | K | K | | µA | nm²/ps | $10^{47} \cdot m^{-3} \cdot J^{-1}$ | meV | nm | nm | µm |
| 1 | 53.4 | 73.9 | 6 | 12 | 0.16 | 445 | 12.73 | 10.85 | 9.3 | 14.5 | 48.7 | 3.6 | 2.3 | 4.0 | 403.5 | 65.1 |
| 2 | 82.9 | 145.1 | 6 | 26 | 0.51 | 393 | 12.37 | 10.63 | 9.9 | 24.4 | 52.8 | 3.7 | 2.2 | 4.2 | 384.7 | 59.2 |
| 3 | 170.6 | 141.4 | 6 | 12 | 0.23 | 431 | 12.63 | 10.72 | 9.8 | 54.5 | 54.4 | 3.3 | 2.3 | 4.3 | 399.0 | 63.7 |

Table 2

| Sample # | $T$ | $U_{VAP,m}/k_B$ | $\varepsilon$ | $I_B$ | $E_B/k_B$ | $G_{B,max}/k_B$ | $x_B$ | $T_{co}$ | $\Delta F_{PS}/k_B$ | $\alpha_{VAP,m}$ | $\alpha_{VH}$ | $\alpha_{VT}$ | $\alpha_{TPS}$ | $\alpha_{QPS}$ |
|---|---|---|---|---|---|---|---|---|---|---|---|---|---|---|
| | K | K | | µA | K | K | nm | K | K | Hz | Hz | Hz | Hz | Hz |
| 1 | 5.7 | 3671.5 | 1.38 | 11.6 | 555.0 | 1027.3 | 46.0 | 0.6 | 2289.7 | 2.59E+24 | 3.77E+34 | 1.86E+140 | 3.00E+55 | 4.88E+35 |
| 2 | 5.7 | 4455.8 | 1.24 | 12.7 | 605.2 | 1374.5 | 75.4 | 0.3 | 3787.3 | 1.30E+28 | 2.56E+37 | 2.69E+197 | 2.05E+85 | 1.93E+60 |
| 3 | 4.5 | 4699.5 | 1.10 | 11.9 | 566.3 | 1707.3 | 163.1 | 0.1 | 7475.2 | 8.35E+35 | 9.83E+42 | 5.50E+206 | 3.02E+224 | 7.18E+256 |

Figure captions

Figure 1. Electrograph of the nanoscale meander with the smallest conduction path. After electron-beam lithography the structure was protected, while a reactive plasma etched 8 nm into a 10 nm high NbN film deposited onto the $Al_2O_3$ substrate. The remaining 2 nm NbN ensures the drain of the electrons from the scanning beam during electrography and therefore eliminates charging effects during the scan. $b_I = 2a_I$ was set for scaling purposes.

Figure 2. Transition into superconductivity measured in transport measurements on a logarithmic scale. Resistance data above the phase transition can be well described by fluctuation conductivity (Eq. (1), red curve). The inset shows an appropriate representation of the data to extract $T_c(0)$ from the resistivity measurements [68]. Below $T_c(0)$ the resistance data follow the expectations for BKT phase transition of VAPs over many orders of magnitude in resistivity (Eq. (12), blue curve).

Figure 3. Temperature dependence of the upper critical field. The solid lines represent linear approximations of the experimental data as described in the text.

Figure 4. Measured fluctuation rates (see chapter III) and their description within the theoretical models discussed in chapter IV. The data are plotted on a logarithmic scale vs. the normalized bias-current $I_b$.

Figure 5. Gibbs free energy barrier for vortex entry, plotted against the relative coordinate perpendicular to the conduction path, for all three investigated samples. For the nanoscale meander with the narrowest conduction path (sample 1), the dependence on the applied

current is also shown. The barrier widths $x_B$ relevant for quantum mechanical tunneling of vortices are also indicated.

Figure 6. Excitation energies $E_{exc}$ in units of $k_B$ for the different models describing the fluctuation rates, calculated for the sample with the narrowest conduction path. For wider conduction paths, the energy scales are larger (not shown).

Figure 7. Normalized fluctuation rates as calculated within the theoretical framework of chapter III. The quantum and thermally activated phase slip models show a current dependence of the normalized fluctuation rate that drops too quickly with decreasing current for explaining our measured data shown in Fig. 4.

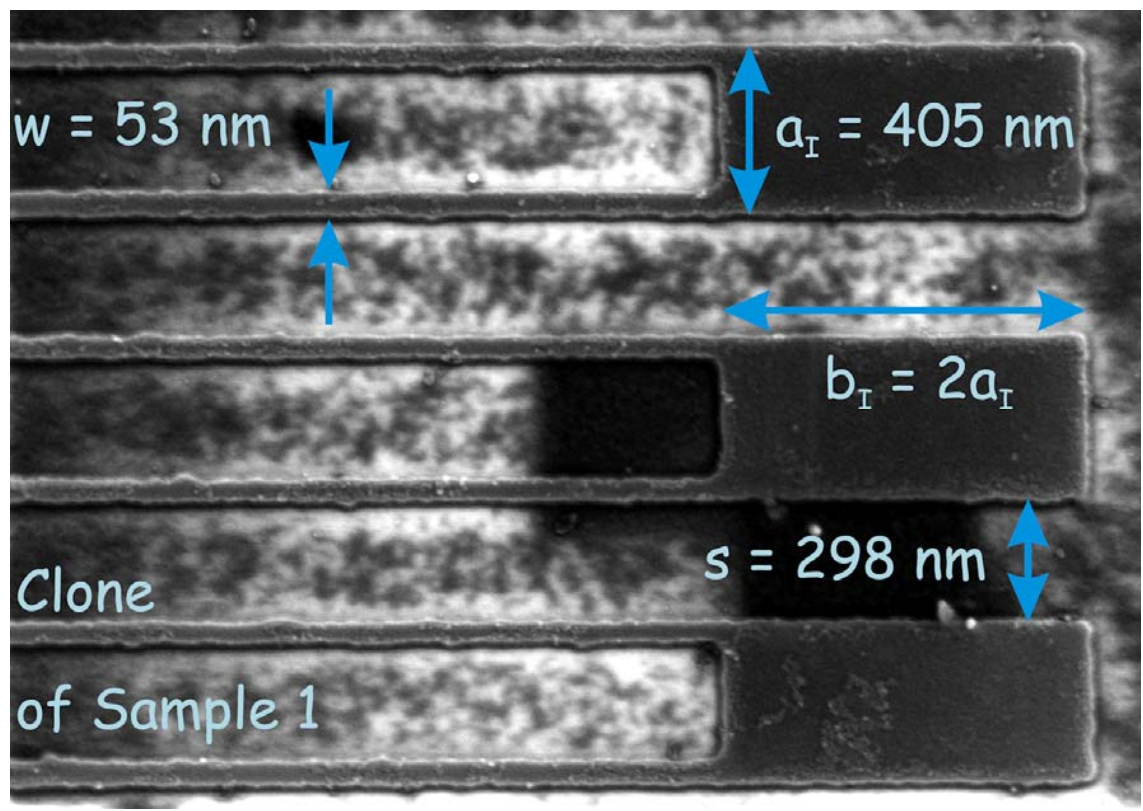

Fig. 1

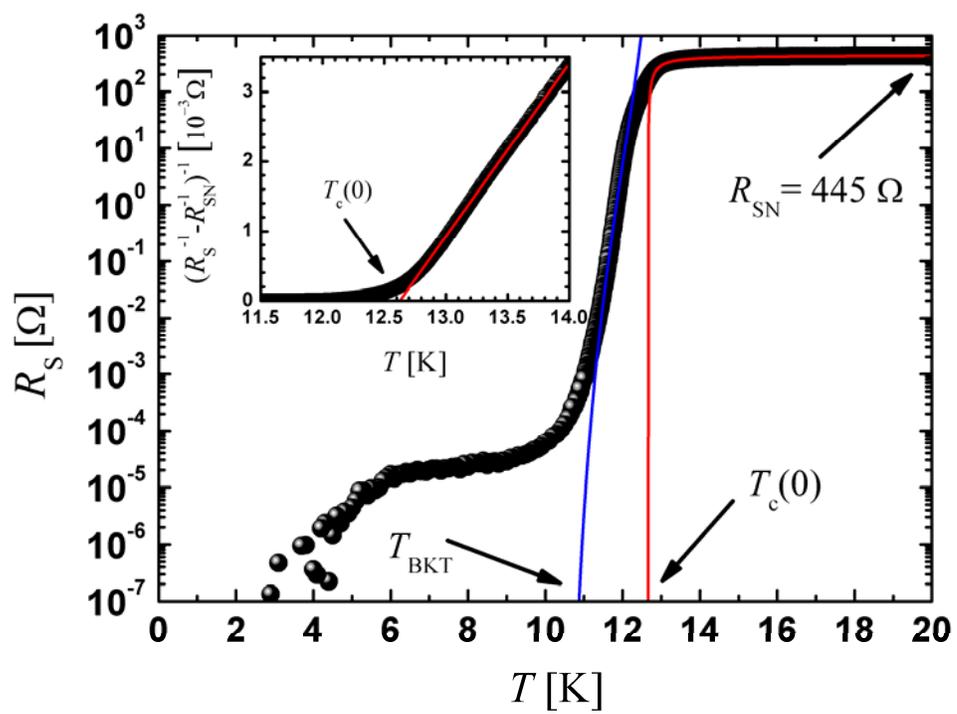

Fig. 2

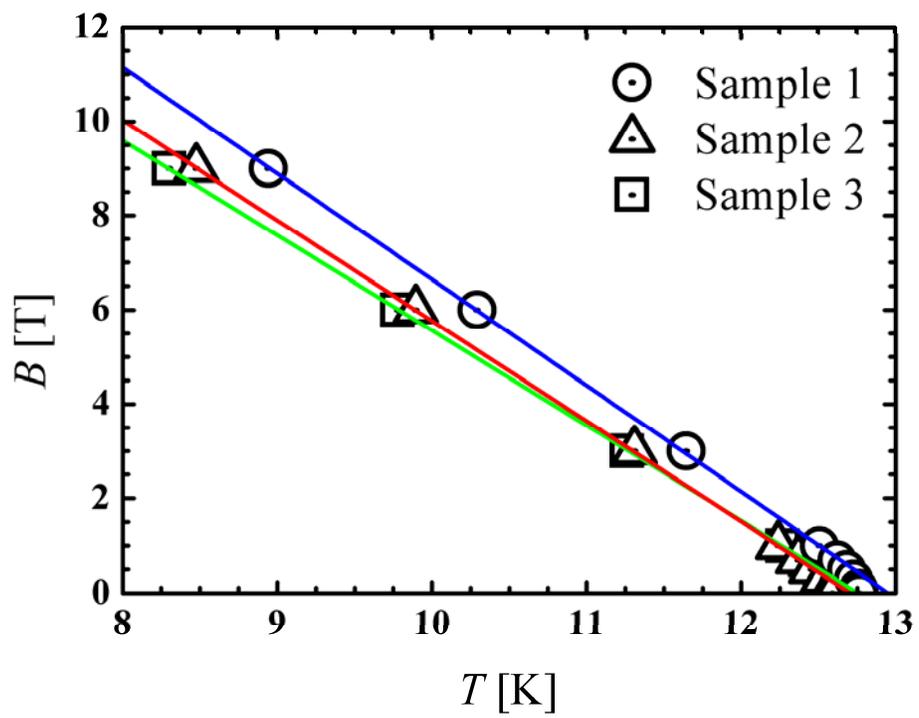

Fig. 3

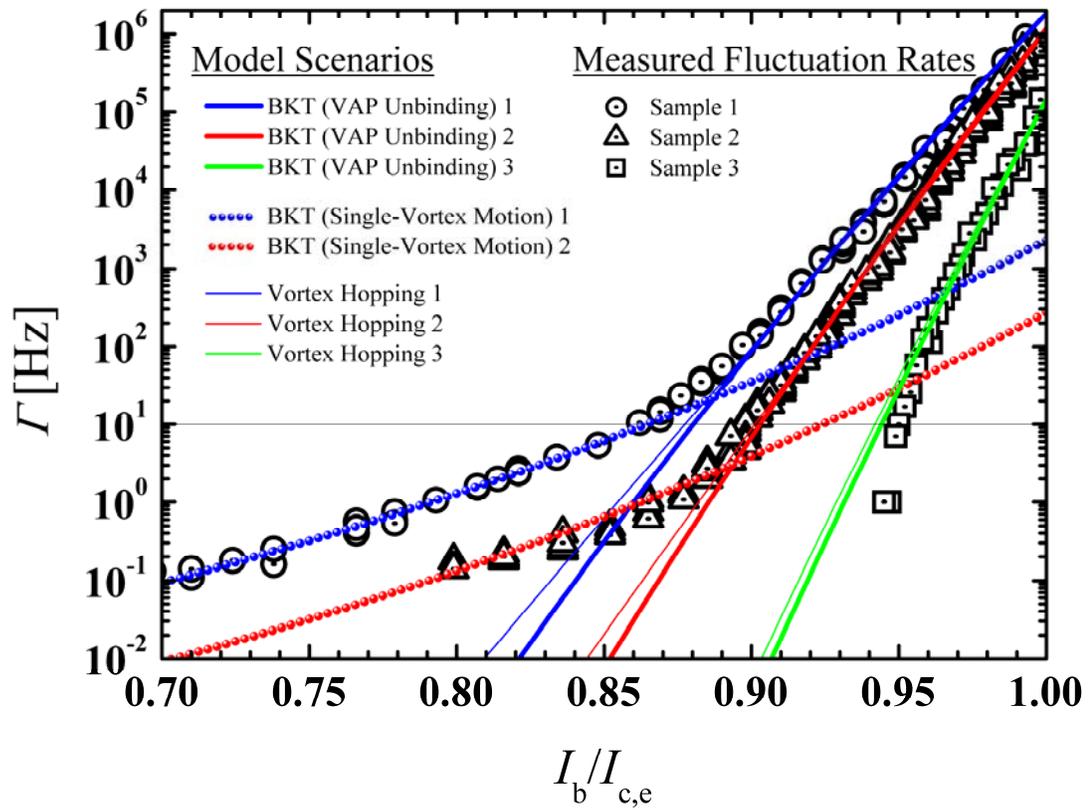

Fig. 4

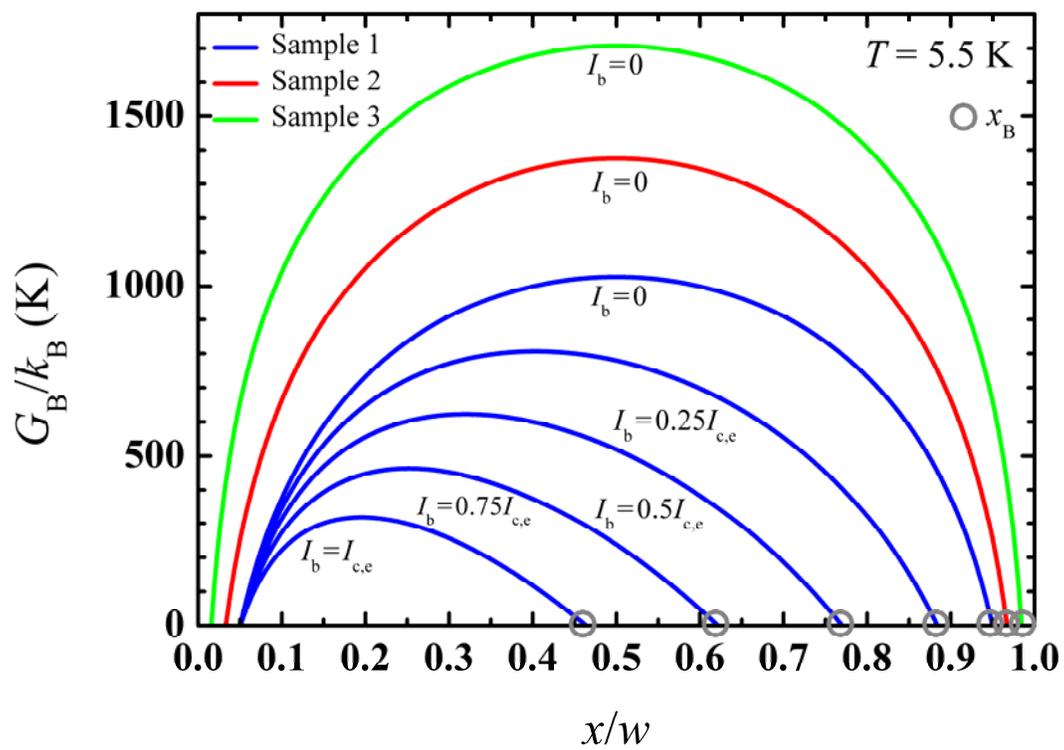

Fig. 5

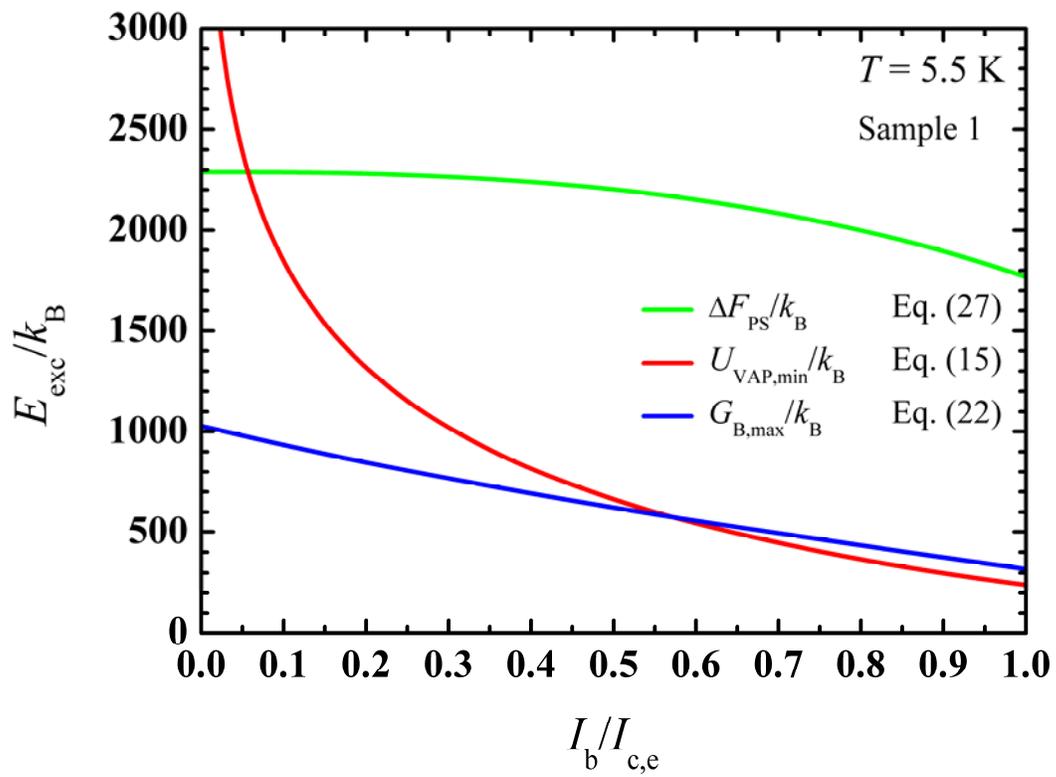

Fig. 6

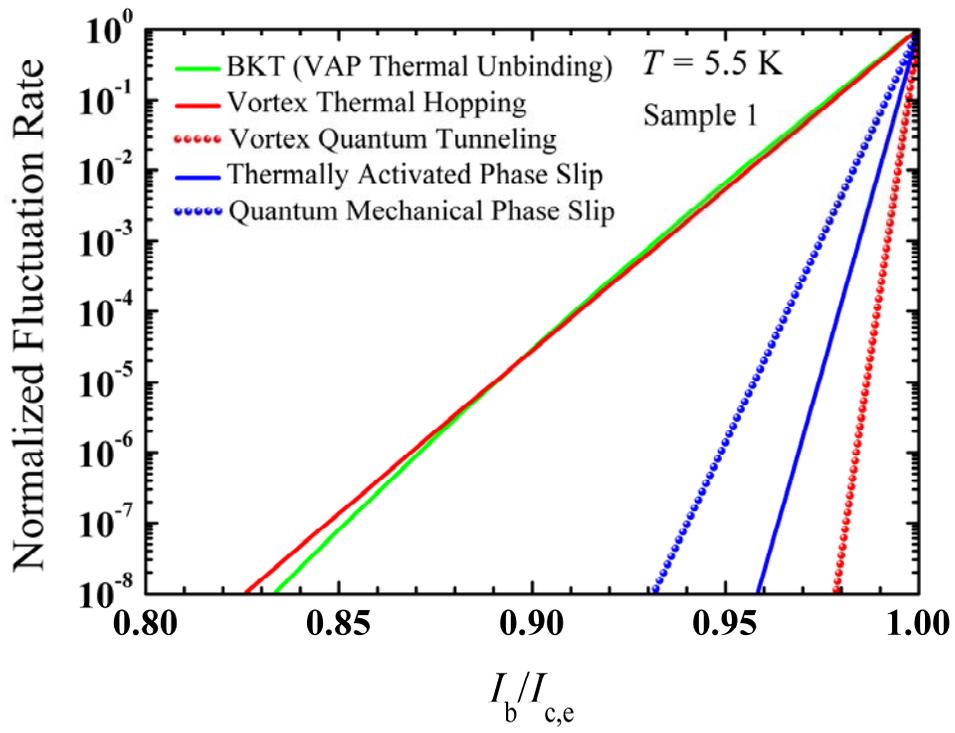

Fig. 7